\documentstyle[aps,12pt]{revtex}
\newcommand{\bb}{\begin{equation}}
\newcommand{\ee}{\end{equation}}
\newcommand{\ba}{\begin{array}}
\newcommand{\ea}{\end{array}}

\begin {document}
\baselineskip 2.2pc

\title{Charged particles in a rotating magnetic field\thanks
{published in Phys. Rev. A {\bf 63} (2001) 012108.}}
\author{Qiong-gui Lin\thanks{E-mail addresses: 
        qg\_lin@163.net, qg\_lin@263.net}}
\address{China Center of Advanced Science and Technology (World
    Laboratory),\\
        P.O.Box 8730, Beijing 100080, People's Republic of China
        \thanks{not for correspondence}\\
        and\\
        Department of Physics, Zhongshan University, Guangzhou
        510275,\\
        People's  Republic of China}

\maketitle
\vfill

\begin{abstract}
\baselineskip 15pt
{\normalsize We study the valence electron of an alkaline atom or
a general charged particle with arbitrary spin and with magnetic
moment moving in a rotating magnetic field.
By using a time-dependent unitary transformation,
the Schr\"odinger equation with the time-dependent Hamiltonian can
be reduced to a Schr\"odinger-like equation with a time-independent
effective Hamiltonian. 
Eigenstates of the effective Hamiltonian correspond to cyclic
solutions of the original Schr\"odinger equation. The nonadiabatic
geometric phase of a cyclic solution can be expressed in terms of the
expectation value of the component of the total angular momentum along
the rotating axis, regardless of whether the solution is explicitly
available. For the alkaline atomic electron and a strong magnetic
field, the eigenvalue problem of the effective Hamiltonian is
completely solved, and the geometric phase turns out to be a linear
combination of two solid angles. For a weak magnetic field, the same
problem is solved partly. 
For a general charged particle, the problem is solved
approximately in a slowly rotating magnetic field,
and the geometric phases are also calculated.}
\end{abstract}
\vfill
\leftline {PACS number(s): 03.65.Ta}
\newpage
\baselineskip 15pt

\section{Introduction}               %I

In quantum mechanics, the Schr\"odinger equation, even with a
time-independent Hamiltonian, can be solved analytically only
in a few cases. With a time-dependent Hamiltonian, the problem
is more difficult and less examples are well studied. One of the
well-studied examples is a nonrelativistic neutral particle with spin
and magnetic moment in a rotating magnetic field [1-5].
This simple example has
received much attention because of its relevance to the problem of
geometric phases [6-8]. As an ordinary physical problem, it may be of
equal interest to consider a charged particle in a
rotating magnetic field. It appears to us that this problem was not
considered in the literature. Because the problem can be treated
analytically and some exact solutions are available, we
will deal with it in this paper.

We will consider an electron moving in a central potential plus a
strong rotating magnetic field in Sec. II. This can describe the
valence eletron of an alkaline atom or the one of a hydrogen atom
under the influence of the external magnetic field. In Sec.
III a weak rotating magnetic field is considered. In Sec. IV we
consider a general charged particle with arbitrary spin and with
magnetic moment moving in a rotating magnetic field, without the
central potential. The Hamiltonians for all these systems are
time dependent. We use a time-dependent unitary transformation to
reduce the Schr\"odinger equation to a Schr\"odinger-like one
with a time-independent effective Hamiltonian. Eigenstates of the
effective Hamiltonian correspond to
cyclic solutions of the original Schr\"odinger equation. The
nonadiabatic geometric phase of a cyclic solution can be expressed in
terms of the expectation value of the component of the total angular
momentum along the rotating axis, regardless of whether the solution
is explicitly available. For the valence electron and a strong magnetic
field considered in Sec. II, the
eigenvalue problem of the effective Hamiltonian can be completely
solved, and the geometric phase is obtained as a linear combination of
two solid angles, of which one is subtended by the trace of the
orbit angular momentum and the other by that of the spin.
For a weak magnetic field in Sec. III, the spin-orbit
coupling has to be included. This makes the effective Hamiltonian more
complicated, and the eigenvalue problem can only be solved partly.
In Sec. IV a scalar potential is absent but a quadratic term of the
magnetic field cannot be discarded. Thus the problem in this section
is rather different. For a slowly rotating magnetic field, one of the
terms in the effective Hamiltonian can be treated as a small
perturbation and the engenvalue problem of the remaining terms can be
solved exactly. In this approximation, the geometric phases of the
cyclic solutions can be calculated explicitly and the results take a
form similar to that in Sec. II. We give a brief summary
and some discussions in Sec. V.

\section{Alkaline atomic electron in a strong
 rotating magnetic field}   %II

Consider the valence electron of an alkaline atom or the electron in
a hydrogen atom. Impose to the system a magnetic field ${\bf B}(t)$
that has a constant magnitude $B$ and rotates about some fixed axis
at a constant angle $\theta_B$ and with a constant frequency $\omega$. 
The rotating axis is chosen as the $z$ axis of the coordinate system.
So the magnetic field is
\bb
{\bf B}(t)=B{\bf n}(t), \quad
{\bf n}(t)=(\sin\theta_B\cos\omega t,\sin\theta_B\sin\omega t,
\cos\theta_B).
\ee     %1
We take $B>0$ without loss of generality.
The motion of the electron in the central potential of the nucleus
(and the other electrons in the inner shells for alkaline atoms)
and the above rotating magnetic field
may be described by the Schr\"odinger equation
$$
i\hbar\partial_t\psi=H(t)\psi,
\eqno(2{\rm a})$$
where
$$
H(t)=H_0+\mu_{\rm B}B({\bf l}+2{\bf s})\cdot{\bf n}(t),
\eqno(2{\rm b})$$
where
$$
H_0={{\bf p}^2\over 2M}+V(r)
\eqno(2{\rm c})$$
is the Hamiltonian of the electron when the external
magnetic field is absent, $M$ and $\mu_{\rm B}$ are respectively the
reduced mass and the Bohr magneton of the electron, 
${\bf l}={\bf r}\times{\bf p}/\hbar$ is the orbit angular
momentum operator in unit of $\hbar$, and {\bf s} the spin in the
same unit. Let us make some remarks on the above Hamiltonian $H(t)$.
First, we are considering a strong magnetic field, so a spin-orbit
coupling term is omitted. Second, the magnetic field is not too strong
such that a quadratic term in ${\bf B}(t)$ can be omitted as well.
Omission of this term also implies that we would not consider very
high excited states, especially scattering states, because the term
is also quadratic in $r$. Third, the rotating magnetic field would
generate a time varying electric field. Thus we are indeed dealing
with a time varying electromagnetic field.
In the nonrelativistic limit,
however, the electric field does not enter the Schr\"odinger equation.
What enters the equation is the vector potential $A_\mu$, which has
been taken as
\addtocounter{equation}{1}
\bb
{\bf A}(t)=\textstyle{\frac 12}{\bf B}(t)\times{\bf r},
\ee     %3
and $-eA_0=V(r)$, where we denote the charge of the electron as $-e$.
The vector
potential ${\bf A}(t)$ produces the magnetic field (1) and results
in the Eq. (2). It also produces the above mentioned electric field.
When Eq. (2) is derived from the relativistic Dirac equation, a term
of the form ${\bbox\alpha}\cdot{\bf E}$ has been discarded, where
${\bbox\alpha}=\gamma^0{\bbox\gamma}$ and the $\gamma$'s are the
Dirac matrices. In this term {\bf E} contains two parts, one from
$V(r)$ and the other from ${\bf A}(t)$. The contribution to the
Hamiltonian from the first part can be neglected in comparison with
$V(r)$, as far as bound states are concerned. The contribution from
the second
part can be neglected in comparison with the second term in Eq. (2b)
for bound states if $\omega\ll 10^{18}$ Hz (by a rough estimation)
which imposes no practical restriction on $\omega$. Therefore one
can indeed forget the existence of the time varying electric field for
the situation being considered.

Now we are going to solve the equation (2). We make a time-dependent
unitary transformation
\bb
\psi(t)=W(t)\Phi(t),
\ee     %4
where
\bb
W(t)=\exp(-i\omega t j_z),
\ee     %5
where $j_z$ is the $z$-component of the total angular momentum
(in unit of $\hbar$)
\bb
{\bf j=l+s}.
\ee     %6
This transformation is a generalization of the one used in solving
the Schr\"odinger equation for a neutral particle with arbitrary
spin in the rotating magnetic field [2]. It is not difficult to show
that
\bb
W^\dagger(t)H(t)W(t)=H(0),
\ee     %7
where
\bb
H(0)=H_0+\mu_{\rm B}B({\bf l}+2{\bf s})\cdot{\bf n}_0,
\ee     %8
where ${\bf n}_0={\bf n}(0)=(\sin\theta_B,0,\cos\theta_B)$.
Thus we obtain the following equation for $\Phi$:
\bb
i\hbar\partial_t\Phi=H_{\rm eff}\Phi,
\ee     %9
where the effective Hamiltonian
\bb
H_{\rm eff}=H(0)-\hbar\omega j_z=H_0+\hbar\omega_0({\bf l}+2{\bf s})
\cdot{\bf n}_0-\hbar\omega(l_z+s_z)
\ee     %10
is time independent. In the above equation we have define $\omega_0=
\mu_{\rm B}B/\hbar$ which is positive. Now let us define two new
frequencies
\bb
\omega_L=(\omega_0^2+\omega^2-2\omega_0\omega\cos\theta_B)^{1/2},
\ee     %11
\bb
\omega_S=(4\omega_0^2+\omega^2-4\omega_0\omega\cos\theta_B)^{1/2},
\ee     %12
and introduce two unit vectors
\bb
{\bf n}_L=(\sin\theta_L, 0, \cos\theta_L),
\ee     %13
\bb
{\bf n}_S=(\sin\theta_S, 0, \cos\theta_S),
\ee     %14
where
\bb
\sin\theta_L={\omega_0\sin\theta_B\over\omega_L},\quad
\cos\theta_L={\omega_0\cos\theta_B-\omega\over\omega_L},
\ee     %15
\bb
\sin\theta_S={2\omega_0\sin\theta_B\over\omega_S},\quad
\cos\theta_S={2\omega_0\cos\theta_B-\omega\over\omega_S}.
\ee     %16
Then the effective Hamiltonian can be written as
\bb
H_{\rm eff}=H_0+\hbar\omega_L{\bf l}\cdot{\bf n}_L +\hbar\omega_S
{\bf s}\cdot{\bf n}_S.
\ee     %17
Therefore the effective direction of the magnetic field is different
for the orbit and spin angular momentum. The eigenvalue problem of
this effective Hamiltonian can be completely solved. This can be
easily realized when it is recasted in the following form.
\bb
H_{\rm eff}=\exp(-i\theta_L l_y-i\theta_S s_y)(H_0+\hbar\omega_L l_z
+\hbar\omega_S s_z)\exp(i\theta_L l_y+i\theta_S s_y).
\ee     %18
We denote the common eigenstates of $(H_0,{\bf l}^2,l_z)$ as
$\zeta_{nlm}^0$ and the eigenstates of $s_z$ as $\chi^0_{m_s}$,
the eigenvalues are respectively $(\epsilon_{nl}, l(l+1), m)$
and $m_s$, where $\epsilon_{nl}$ are the energy spectrum of the
electron in the absence of the magnetic field. The eigenstates of 
$H_{\rm eff}$ are then
\bb
\varphi_{nlmm_s}=\zeta_{nlm}\chi_{m_s},
\ee     %19
where
\bb
\zeta_{nlm}=\exp(-i\theta_L l_y)\zeta_{nlm}^0,\quad
\chi_{m_s}=\exp(-i\theta_S s_y)\chi_{m_s}^0.
\ee     %20
The corresponding energy eigenvalues are
\bb
E_{nlmm_s}=\epsilon_{nl}+m\hbar\omega_L+m_s\hbar\omega_S.
\ee     %21
There is no degeneracy in the quantum numbers. This is different from
the case in a static magnetic field (Paschen-Back effect) where some
degeneracy is preserved. Unfortunately, the above energy levels are
not observable since $H_{\rm eff}$ is not a physical quantity. Note
that $\varphi_{nlmm_s}$ are also the common eigenstates of the
operators $(H_0,{\bf l}^2,{\bf l}\cdot{\bf n}_L,{\bf s}\cdot
{\bf n}_S)$ with eigenvalues $(\epsilon_{nl}, l(l+1), m, m_s)$.
Since $H_{\rm eff}$ is time-independent, Eq. (9) has the formal
solution
\bb
\Phi(t)=U_{\rm eff}(t)\Phi(0),\quad
U_{\rm eff}(t)=\exp(-iH_{\rm eff}t/\hbar).
\ee     %22
With the obvious relation $\psi(0)=\Phi(0)$, the time-dependent
Schr\"odinger equation (2) is formally solved as
\bb
\psi(t)=U(t)\psi(0),\quad
U(t)=W(t)U_{\rm eff}(t).
\ee     %23
Since $U(t)$ involves no chronological product, this solution is
convenient for practical calculations.

Now we show that eigenstates of the effective Hamiltonian correspond
to cyclic solutions of Eq. (2). We take the initial condition
\bb
\psi_i(0)=\varphi_i=\varphi_{nlmm_s},
\ee     %24
and calculate $\psi_i(T)$ where $T=2\pi/\omega$ is the period of the
rotating magnetic field. Here for convenience we use one subscript $i$
to represent all the quantum numbers $nlmm_s$. Obviously,
$U_{\rm eff}(t)\varphi_i=\exp(-iE_{i}t/\hbar)\varphi_i$, valid for all
$t$, and $W(T)\varphi_i=\exp(-i2\pi l_z)\zeta_{nlm}\exp(-i2\pi s_z)
\chi_{m_s}$. Because
\bb
\zeta_{nlm}=\sum_{m'}D^l_{m'm}(0,\theta_L,0)\zeta_{nlm'}^0,\quad
\chi_{m_s}=\sum_{m'_s}D^{1/2}_{m'_s m_s}(0,\theta_S,0)\chi_{m'_s}^0,
\ee     %25
where the $D$'s are Wigner functions, we have $W(T)\varphi_i=
\exp(-i2\pi m-i2\pi m_s)\varphi_i$. Finally we obtain
\bb
\psi_i(T)=\exp(-iE_{i}T/\hbar-i2\pi m-i2\pi m_s)\psi_i(0).
\ee     %26
Hence it is indeed a cyclic solution, and the total phase change
in a period is
\bb
\delta_i=-E_{i}T/\hbar-2\pi m-2\pi m_s
=-E_{i}T/\hbar-\pi,\quad {\rm mod} ~2\pi.
\ee     %27
To determine the dynamic phase, we should calculate
$$
\langle H(t)\rangle_i\equiv (\psi_i(t), H(t)\psi_i(t))=(\psi_i(0),
W^\dagger H(t)W\psi_i(0))=(\psi_i(0), H(0)\psi_i(0)).
$$
Because $H(0)=H_{\rm eff}+\hbar\omega j_z$, we have
\bb
\langle H(t)\rangle_i=E_i+\hbar\omega\langle j_z\rangle_i.
\ee     %28
Here $\langle j_z\rangle_i=(\psi_i(0), j_z\psi_i(0))=
(\psi_i(t), j_z\psi_i(t))$ is the expectation value of $j_z$ in the
state $\psi_i(t)$, and it is time independent. Note that
$\langle H(t)\rangle_i$ is also independent of $t$. Thus the state
$\psi_i(t)$ is somewhat similar to a stationary state in a system
with a time-independent Hamiltonian. The dynamic phase is
\bb
\beta_i=-\hbar^{-1}\int_0^T dt\;\langle H(t)\rangle_i=
-E_iT/\hbar-2\pi\langle j_z\rangle_i.
\ee     %29
Therefore the nonadiabatic geometric phase is
\bb
\gamma_i=\delta_i-\beta_i=-\pi+2\pi\langle j_z\rangle_i,
\quad {\rm mod} ~2\pi,
\ee     %30
and is determined by $\langle j_z\rangle_i$. This is a generalization
of the result for a neutral particle [2]. For the above calculations
to stand, only two facts are necessary. First, the initial state is an
eigenstate of
$H_{\rm eff}$. Second, it can be expanded as a linear combination of
the common eigenstates of $(l_z, s_z)$. The latter is independent of
the above specific form of $H_{\rm eff}$. Therefore the result
(30) is valid regardless of whether $\varphi_i$ is explicitly
available or not, and
is convenient for approximate calculations if necessary.

For the present case, the above geometric phase can be worked out
explicitly. Indeed, it is easy to show that
$$
\exp(i\theta_L l_y)l_z\exp(-i\theta_L l_y)
=-\sin\theta_L l_x+\cos\theta_L l_z,
$$
thus in the $i$th cyclic solution with $\psi_i(0)=\varphi_i$,
we have
\bb
\langle l_z\rangle_i=(\varphi_i, l_z\varphi_i)
=(\zeta_{nlm}, l_z\zeta_{nlm})=(\zeta_{nlm}^0,\exp(i\theta_L l_y)l_z
\exp(-i\theta_L l_y)\zeta_{nlm}^0)=m\cos\theta_L.
\ee     %31
Similarly, we have
\bb
\langle s_z\rangle_i=m_s\cos\theta_S.
\ee     %32
Therefore the geometric phase takes the form
\bb
\gamma_i=-m\Omega_L-m_s\Omega_S, \quad {\rm mod} ~2\pi,
\ee     %33
where
\bb
\Omega_L=2\pi(1-\cos\theta_L),\quad
\Omega_S=2\pi(1-\cos\theta_S)
\ee     %34
are the solid angles subtended by the traces of orbit and spin angular
momentum. Indeed, it is easy to show that
\bb
(\psi_i(t), {\bf l}\psi_i(t))=m
(\sin\theta_L\cos\omega t,\sin\theta_L\sin\omega t,\cos\theta_L),
\ee     %35
\bb
(\psi_i(t), {\bf s}\psi_i(t))=m_s
(\sin\theta_S\cos\omega t,\sin\theta_S\sin\omega t,\cos\theta_S).
\ee     %36
Therefore the geometric nature of the results is quite obvious. It
should be remarked that the orbit and spin angular momentum both
precess synchronously with the magnetic field, but each at a different
angle with the rotating axis.

In the above paragraph, we have shown that the initial condition
$\psi_i(0)=\varphi_i$ leads to a cyclic solution. On the other hand,
a cyclic solution need not always take such an initial condition.
If the parameters are appropriately chosen, one can find other cyclic
solutions explicitly. For example, if $B$ (or $\omega_0$) and
$\theta_B$ are chosen such that
\bb
\omega_L=N_L\omega,\quad \omega_S=N_S\omega,
\ee     %37
where $N_L$ and $N_S$ are natural numbers, then any solution with the
initial condition
\bb
\psi(0)=\sum_{mm_s}a_{mm_s}\varphi_{nlmm_s}
\ee     %38
is a cyclic solution, where the coefficients $a_{mm_s}$ are arbitrary.
In fact, one can show that
\bb
\psi(T)=\exp[-i\epsilon_{nl}T/\hbar-i(N_S+1)\pi]\psi(0).
\ee     %39
To conclude this section we give a specific realization of the
condition (37). If one chooses $\omega_0/\omega=\sqrt{3/2}$ and
$\cos\theta_B=\sqrt 3/2\sqrt 2$, then the condition (37) holds
with $N_L=1$, $N_S=2$.

\section{Alkaline atomic electron in a weak
 rotating magnetic field}   %III

In this section we consider the same problem as in Sec. II but with
a weak rotating magnetic field. In this case a spin-orbit coupling
term has to be included in the Hamiltonian, and the Schr\"odinger
equation reads
$$
i\hbar\partial_t\psi=H(t)\psi,
\eqno(40{\rm a})$$
where
$$
H(t)=H_0+\mu_{\rm B}B({\bf l}+2{\bf s})\cdot{\bf n}(t)
+\hbar^2\xi(r) {\bf l}\cdot{\bf s}.
\eqno(40{\rm b})$$
We make use of the time-dependent unitary transformation (4-5). Since
the spin-orbit coupling term is invariant under the transformation,
Eq. (7) holds for the present case. Thus we get for $\Phi$ the
equation
\addtocounter{equation}{1}
\bb
i\hbar\partial_t\Phi=H_{\rm eff}\Phi,
\ee     %41
where the effective Hamiltonian is now given by
\bb
H_{\rm eff}=H_0+\hbar\omega_L{\bf l}\cdot{\bf n}_L +\hbar\omega_S
{\bf s}\cdot{\bf n}_S+\hbar^2\xi(r){\bf l}\cdot{\bf s}.
\ee     %42
As in Sec. II, $H_{\rm eff}$ is time independent, so we have the
formal solutions for Eqs. (41) and (40)
\bb
\Phi(t)=U_{\rm eff}(t)\Phi(0),\quad
U_{\rm eff}(t)=\exp(-iH_{\rm eff}t/\hbar),
\ee     %43
\bb
\psi(t)=U(t)\psi(0),\quad
U(t)=W(t)U_{\rm eff}(t).
\ee     %44
Because $H_{\rm eff}$ is an Hermitian operator, one can in principle
find a complete set of eigenstates $\{\varphi_i\}$, with eigenvalues
$\{E_i\}$. A cyclic solution of the Schr\"odinger equation can be
obtained by taking anyone of these eigenstates as an initial state,
i.e., $\psi_i(0)=\varphi_i$. In fact,
one can always expand $\varphi_i$ as a
linear combination of the common eigenstates of $(l_z, s_z)$, and
obtain
\bb
\psi_i(T)=\exp(-iE_{i}T/\hbar-i\pi)\psi_i(0).
\ee     %45
The geometric phase of this solution in a period can be calculated in
a way similar to that in Sec. II. The result has the same form:
\bb
\gamma_i=-\pi+2\pi\langle j_z\rangle_i,
\quad {\rm mod} ~2\pi.
\ee     %46

Unfortunately, in the present case the eigenvalue problem of
$H_{\rm eff}$ is difficult to solve in general.
The only operator that commute with
$H_{\rm eff}$ and thus can have common eigenstates is ${\bf l}^2$.
(Of course ${\bf s}^2$ is another such operator, but it is trivial.)
We may denote these common eigenstates as $\varphi_i=\varphi_{lp}$
where $p$ represents all other quantum numbers. Here we deal only with
the special case $l=0$ where the eigenstates are explicitly available.
They read
\bb
\varphi_{0p}=\zeta_{n00}^0\exp(-i\theta_S s_y)\chi_{m_s}^0,
\ee     %47
where $\zeta_{n00}^0$ and $\chi_{m_s}^0$ are those used in Sec. II.
In this case the geometric phase can be easily found to be
\bb
\gamma_{0p}=-m_s\Omega_S, \quad {\rm mod} ~2\pi.
\ee     %48

\section{General charged particles in a rotating magnetic field} %III

In this section we consider a general charged particle with an
arbitrary spin $s$ ($s=1/2,1,3/2,\ldots$)
and with magnetic moment moving in the rotating magnetic field (1),
without a central potential. We denote the charge of the particle by
$q$, the mass by $M$, the spin angular momentum by {\bf s},
and the magnetic
moment by $\mu$. In this case the Schr\"odinger equation reads
$$
i\hbar\partial_t\psi=H(t)\psi,
\eqno(49{\rm a})$$
where
$$
H(t)={1 \over 2M}\left[{\bf p}-{q\over c}{\bf A}(t)\right]^2
-{\bbox\mu}\cdot{\bf B}(t),
\eqno(49{\rm b})$$
where ${\bbox\mu}=\mu{\bf s}/s$, and ${\bf A}(t)$ is given by Eq. (3),
which produces the magnetic field. Note that ${\bf A}(t)$
also produces an
time-dependent electric field as pointed out before. Now that the
charged particle is not necessarily confined in a small region, there
seems no reason to neglect the existence of the electric field. Thus
we are actually dealing with an electromagnetic field. 
Fortunately, the electric field does not
enter the equation as long as we confine ourselves to a
nonrelativistic theory. This is also the reason why we
only speak about the magnetic field.

The above Hamiltonian can be rewritten in the form
\addtocounter{equation}{1}
\bb
H(t)={{\bf p}^2\over 2M}+{\frac 12}M\omega_1^2[r^2-({\bf r}\cdot
{\bf n}(t))^2]-\epsilon(q)\hbar\omega_1{\bf l}\cdot{\bf n}(t)
-\epsilon(\mu)\hbar\omega_2{\bf s}\cdot{\bf n}(t),
\ee     %50
where $\omega_1=|q|B/2Mc$, $\omega_2=|\mu|B/s\hbar$, both being
positive, and $\epsilon(q)$, $\epsilon(\mu)$ are sign functions.
As before, we make use of the time-dependent unitary transformation
(4-5). It can be shown that $W^\dagger(t)H(t)W(t)=H(0)$ still holds
in the present case.
Thus we obtain the following equation for $\Phi$:
\bb
i\hbar\partial_t\Phi=H_{\rm eff}\Phi,
\ee     %51
where 
$$
H_{\rm eff}=H(0)-\hbar\omega j_z,
$$
or
\bb
H_{\rm eff}={{\bf p}^2\over 2M}+{\frac 12}M\omega_1^2[r^2-({\bf r}
\cdot{\bf n}_0)^2]-\epsilon(q)\hbar\omega_1{\bf l}\cdot{\bf n}_0
-\epsilon(\mu)\hbar\omega_2{\bf s}\cdot{\bf n}_0-\hbar\omega(l_z+s_z).
\ee     %52
As before, $H_{\rm eff}$ is time independent, so we have the formal
solutions for Eqs. (51) and (49)
\bb
\Phi(t)=U_{\rm eff}(t)\Phi(0),\quad
U_{\rm eff}(t)=\exp(-iH_{\rm eff}t/\hbar),
\ee     %53
\bb
\psi(t)=U(t)\psi(0),\quad
U(t)=W(t)U_{\rm eff}(t).
\ee     %54
Because $H_{\rm eff}$ is an Hermitian operator, one can in principle
find a complete set of eigenstates $\{\varphi_i\}$, with eigenvalues
$\{E_i\}$. A cyclic solution of the Schr\"odinger equation can be
obtained by taking anyone of these eigenstates as an initial state,
i.e., $\psi_i(0)=\varphi_i$. In fact, one can show that
\bb
\psi_i(T)=\exp(-iE_{i}T/\hbar-i2\pi s)\psi_i(0).
\ee     %55
The geometric phase of this solution in a period can be found to be
\bb
\gamma_i=-2\pi s+2\pi\langle j_z\rangle_i,
\quad {\rm mod} ~2\pi.
\ee     %56
These are all familiar results. The remaining task is to find the
eigenvalues and eigenstates of $H_{\rm eff}$, and to work out the
geometric phase explicitly.

Since the effective Hamiltonian is somewhat complicated, we have to
make some approximation. We assume that $\omega\ll\omega_1$,
$\omega\ll\omega_2$, that is, we are considering slowly rotating
magnetic field. This is in fact a rather loose restriction. To see
this, let us make some rough estimation.  For a
magnetic field of the magnitude 1 tesla (which is not very large),
we have for an electron $\omega_1\sim 10^{10}$ Hz, $\omega_2=2
\omega_1$, and for a proton $\omega_1\sim 10^{7}$ Hz, $\omega_2\sim
10^{8}$ Hz. An experimentally available rotating frequency must be
much smaller than these ones. Under this assumption, the term
$\hbar\omega l_z$ in Eq. (52)
is much smaller than the two preceding terms
and thus can be treated as a small perturbation. On the other hand,
the term $\hbar\omega s_z$, which is of the same order of magnitude,
need not be treated approximately. We thus decompose $H_{\rm eff}$
as
\bb
H_{\rm eff}=H^0_{\rm eff}+H'_{\rm eff},
\ee     %57
where 
\bb
H^0_{\rm eff}={{\bf p}^2\over 2M}+{\frac 12}M\omega_1^2[r^2-({\bf r}
\cdot{\bf n}_0)^2]-\epsilon(q)\hbar\omega_1{\bf l}\cdot{\bf n}_0
-\epsilon(\mu)\hbar\omega_S{\bf s}\cdot{\bf n}_S,
\ee     %58
whoes eigenvalue problem will be solved exactly, and
\bb
H'_{\rm eff}=-\hbar\omega l_z
\ee     %59
will be treated as a small perturbation. In Eq. (58),
\bb
\omega_S=[\omega_2^2+\omega^2+2\epsilon(\mu)\omega_2\omega
\cos\theta_B]^{1/2},
\ee     %60
\bb
{\bf n}_S=(\sin\theta_S, 0, \cos\theta_S),
\ee     %61
where
\bb
\sin\theta_S={\omega_2\sin\theta_B\over\omega_S},\quad
\cos\theta_S={\omega_2\cos\theta_B+\epsilon(\mu)\omega\over\omega_S}.
\ee     %62
It is not difficult to show that 
\bb
H^0_{\rm eff}=\exp(-i\theta_B l_y-i\theta_S s_y)H^z_{\rm eff}
\exp(i\theta_B l_y+i\theta_S s_y),
\ee     %63
where
\bb
H^z_{\rm eff}={{\bf p}^2\over 2M}+{\frac 12}M\omega_1^2 \rho^2
-\epsilon(q)\hbar\omega_1 l_z-\epsilon(\mu)\hbar\omega_S s_z
\ee     %64
is the Hamiltonian of a charged particle in a static magnetic field
along the $z$ axis (but note that $\omega_2$ is replaced by
$\omega_S$). In the above equation $\rho$ is a cylindrical coordinate.
The eigenvalue problem of $H^z_{\rm eff}$ is well known. In the
cylindrical coordinates, the eigenfunctions are
\bb
u_i=u_{n_{\rho} n_z m m_s}(\rho,\phi,z,s_z)
=N_{n_{\rho} m}\exp\left(-\frac 12\alpha^2\rho^2\right)
(\alpha\rho)^{|m|}L_{n_{\rho}}^{|m|}(\alpha^2\rho^2)
{e^{im\phi}\over\sqrt{2\pi}}{\exp(i2\pi n_z z/d)\over \sqrt d}
\chi^0_{m_s},
\ee     %65
where $m$ and $n_z$ take on integer values, $n_\rho=0,1,2,\ldots$,
$m_s=s,s-1,\ldots,-s$,
$\alpha=\sqrt{M\omega_1/\hbar}$, $d$ is a length in
the $z$ direction for box normalization, the $L_{n_{\rho}}^{|m|}$ are
Laguerre polynomials [9], and the normalization constant
$$
N_{n_{\rho} m}=\alpha\left[{2n_{\rho}!\over
\Gamma(n_{\rho}+|m|+1)}\right]^{1/2}.
$$
Note that these are also commom eigenstates of $(l_z,s_z)$ with
eigenvalues $(m, m_s)$. The eigenvalue of $H^z_{\rm eff}$ is
\bb
E_i^0=E_{n_{\rho} n_z m m_s}^0=(2n_{\rho}+|m|+1)\hbar\omega_1
+{2 n_z^2 \pi^2\hbar^2\over Md^2}-\epsilon(q)m\hbar\omega_1
-\epsilon(\mu)m_s\hbar\omega_S.
\ee      %66
There are some degeneracy in the quantum numbers $m$ and $n_\rho$.
The eigenfunctions of $H^0_{\rm eff}$ are given by
\bb
\varphi_{i}^0=\exp(-i\theta_B l_y-i\theta_S s_y)u_i.
\ee     %67
The corresponding energy eigenvalues are still given by Eq. (66).
We will use these $\varphi_{i}^0$ as the approximate eigenfunctions
of $H_{\rm eff}$. The lowest-order corrections to the energy
eigenvalues are given by the expectation values of $H'_{\rm eff}$ in
these eigenstates. The corrected energy levels are
\bb
E_i=E_{n_{\rho} n_z m m_s}=(2n_{\rho}+|m|+1)\hbar\omega_1
+{2 n_z^2 \pi^2\hbar^2\over Md^2}-m\hbar[\epsilon(q)\omega_1+\omega
\cos\theta_B]-\epsilon(\mu)m_s\hbar\omega_S.
\ee      %68
Now there is no degeneracy in the quantum numbers. The geometric
phase of the $i$th state in a period is approximately
given by
\bb
\gamma_i=-m\Omega_B-m_s\Omega_S, \quad {\rm mod} ~2\pi,
\ee     %69
where $\Omega_S$ has been defined in Eq. (34), and $\Omega_B=2\pi(1-
\cos\theta_B)$. Note that the second term is exact.
The approximation lies in the first term.

Though the above $u_i$ or $\varphi_i^0$ are not eigenfunctions of
${\bf l}^2$, it can still be shown that
\bb
(\psi_i(t), {\bf l}\psi_i(t))=m
(\sin\theta_B\cos\omega t,\sin\theta_B\sin\omega t,\cos\theta_B),
\ee     %70
\bb
(\psi_i(t), {\bf s}\psi_i(t))=m_s
(\sin\theta_S\cos\omega t,\sin\theta_S\sin\omega t,\cos\theta_S).
\ee     %71
Therefore the orbit and spin angular momentum both
precess synchronously with the magnetic field, the former at an 
angle $\theta_B$ (approximate) and the latter at $\theta_S$ (exact)
with the rotating axis. The geometric nature of the result (69) is
thus obvious. 

\section{Summary and discussions}  %V

In this paper we consider a nonrelativistic charged particle moving in
a rotating magnetic field, with or without a central potential.
The case with a central potential may describe the valence electron
in an alkaline atom or the only one in a hydrogen atom.
The Hamiltonian for such a system is time dependent. By making use
of a time-dependent unitary transformation, the Schr\"odinger equation
can be reduced to a Schr\"odinger-like equation with an effective
Hamiltonian which is time independent. In this way one obtains a
formal solution to the original Schr\"odinger equation, which
determines the time evolution of an arbitrary initial state. The
time-evolution operator in this formal solution, unlike the one
for a general time-dependent Hamiltonian, involves no chronological
product, and thus is convenient for practical calculations.
Any solution with one of the eigenstates of the effective Hamiltonian
as an initial state is a cyclic solution. The geometric phase in
a period for such a solution can be expressed in terms of the
expectation value of the component of the total angular momentum
along the rotating axis. This holds regardless of whether the
solution is explicitly available, and is convenient for approximate
calculations whenever necessary. We discuss in detail the alkaline
atomic electron in a strong or weak magnetic field. In a strong
magnetic field everything is worked out explicitly. In a weak field,
however, the effective Hamiltonian is complicated and the eigenvalue
problem of it is explicitly solved only in a special case. For a
general charged particle and without the central potential, the 
eigenvalue problem of the effective Hamiltonian is solved
approximately. It is a rather good approximation if the magnetic
field rotates slowly.

The time-dependent unitary transformation used in this paper is a
generalization of the one previously used for a neutral particle with
arbitrary spin and with magnetic moment in a rotating magnetic
field [2]. It transforms the Schr\"odinger equation to one in a
rotating frame where the magnetic field is static. Thus in that
frame the effective Hamiltonian is time independent. An extended
rotating-frame formalism has been used to study a neutral particle
in a more general time-dependent magnetic field [10]. It should be
remarked, however, that the transformation is merely a mathematical
technique, and the effective equation in the rotating frame does not
describe a real physical problem whose results are observable in that
frame. A real physical problem in a
rotating frame would be much more complicated since the frame is not
an inertial system.

The extension of the present problem to relativistic particles,
either electrically charged or neutral, is currently under
investigation. Progress will be reported subsequently.

\section*{Acknowledgments}

The author is grateful to Professor Guang-jiong Ni for communications
and encouragement. This work was supported by the
National Natural Science Foundation of China.

\end{document}